\documentclass[twocolumn,preprintnumbers,amsmath,amssymb]{revtex4}
\usepackage{graphicx}
\usepackage{dcolumn}
\usepackage{bm}
\usepackage{hyperref} 
\usepackage{multirow} 
\newcommand\beq{\begin{equation}}
\newcommand\eeq{\end{equation}}
\newcommand\bea{\begin{eqnarray}}
\newcommand\eea{\end{eqnarray}}
\usepackage{wrapfig}
\newcommand{\nonum}{\nonumber}
\begin{document}
\title{\bf Decoupled phase of frustrated spin-1/2 antiferromagnetic chains with
and without long range order in the ground state \\}
\author{\bf  Manoranjan Kuma${\rm \bf r^{1}}$ and Z.G. Soo${\rm \bf s^2} $}
\address{\it {\rm  $ ^1 S.$} N. Bose National Centre for Basic Sciences. Block-JD, Sector-III, Kolkata 700098,
India.\\}
\address{\it {\rm  $ ^2 Department$ } of Chemistry, Princeton University, Princeton NJ 08544 \\}
\date{\today}
\begin{abstract}
The quantum phases of one-dimensional spin $s= 1/2$ chains are discussed for models with
two parameters, frustrating exchange $g = J_2 > 0$ between second neighbors and
normalized nonfrustrating power-law exchange with exponent $\alpha$ and distance dependence $r^{-\alpha}$. The ground 
state (GS) at $g = 0$ has long-range order (LRO) for $\alpha < 2$, long-range spin fluctuations 
for $\alpha > 2$. The models conserve total spin $S = S_A + S_B$, have singlet 
GS for any $g$, $\alpha \ge 0$ and decouple at $1/g = 0$ to linear Heisenberg 
antiferromagnets on sublattices $A$ and $B$ of odd and even-numbered sites. 
Exact diagonalization of finite chains gives the sublattice spin $ \langle S^2_A \rangle$, 
the magnetic gap $E_m$ to the lowest triplet state and the excitation $E_{\sigma}$ to the 
lowest singlet with opposite inversion symmetry to the GS. An analytical model that 
conserves sublattice spin has a first order quantum transition at $g_c = 1/4{\rm ln2}$ from 
a GS with perfect LRO to a decoupled phase with $S_A = S_B = 0$ for $g \ge 4/\pi^2$ 
and no correlation between spins in different sublattices. The model with $\alpha = 1$ has 
a first order transition to a decoupled phase that closely resembles the analytical model. 
The bond order wave (BOW) phase and continuous quantum phase transitions 
of finite models with $\alpha \ge 2$  are discussed in terms of GS degeneracy 
where $E_{\sigma}(g) = 0$, excited state degeneracy where $E_{\sigma}(g) = E_m(g)$, 
and $\langle S^2_A \rangle$. The decoupled phase at large frustration 
has nondegenerate GS for any exponent $\alpha$ and excited states related 
to sublattice excitations.
\vskip .4 true cm
\noindent PACS numbers: 75.10.Pq, 75.10.Jm, 64.70.Tg, 73.22.Gk\\ 
\noindent Email:soos@princeton.edu,manoranjan.kumar@bose.res.in 
\end{abstract}
\maketitle

\section{Introduction}
One-dimensional (1D) spin chains have provided a wealth of quantum many-body
problems over the years, starting with Bethe's treatment \cite{r1} of the linear Heisenberg
antiferromagnet (HAF) with exchange $J_1 > 0$ between adjacent $s = 1/2$ sites. The linear
HAF has been extensively studied and generalized \cite{r2,r3,r4}. Second-neighbor exchange 
$J_2 > 0$ is frustrating in chains with either sign of $J_1$. Magnetic frustration has recently been reported in 
copper oxides that contain $s = 1/2$ chains of Cu(II) ions with $J_1 > 0$ \cite{r4p} or $< 0$ \cite{r4pp}. 
Increasing $g  = J_2/J_1$ in the $J_1-J_2$ model generates a continuous quantum 
transition from the HAF ground state to a bond order wave (BOW) phase that has been 
established by multiple theoretical methods \cite{r5,r6,r7} but to the best of our 
knowledge, the BOW phase of spin chains has not been realized experimentally. 
The characterization of the BOW phase is limited: The ground state is doubly 
degenerate, inversion symmetry is broken and there is a finite energy gap $E_m$ 
between the singlet ground state and lowest triplet state. We discuss in this 
paper spin chains with frustrating $J_2 > 0$ and variable-range nonfrustrating 
exchange that narrow and eventually suppress the BOW phase.\\ 

Laflorencie, Affleck and Berciu \cite{r8} studied chains with 
nonfrustrating exchange $J_r(\alpha) \approx (-1)^{(r-1)}/r^{\alpha}$
between spins $p$ and $p + r$. Power laws $\alpha < \alpha_c \approx 2$ 
leads to 1D models whose ground state has long-range order (LRO) while $\alpha > \alpha_c$ 
leads to a spin liquid with long-range spin correlations. There are exact field 
theory results\cite{r9} at $\alpha = 2$. Sandvik \cite{r10} combined frustrating 
second-neighbor exchange $g = J_2 > 0$ with variable-range $J_r(\alpha)$ in spin chains 
that are characterized by two parameters, frustration $g$ and exponent $\alpha$
\begin{eqnarray}
H(g,\alpha)=H_1(\alpha)+gH_2.
\label{eq1}
\end{eqnarray}
$H_2$ is always a linear HAF with unit exchange between neighbors in sublattices A and B of even and odd numbered sites, 
\begin{eqnarray}
H_2=H_A+H_B=\sum_p \vec{s}_p.\vec{s}_{p+2} .
\label{eq2}
\end{eqnarray}
$H_1(\alpha)$ has nonfrustrating exchanges $J_r(\alpha)$ 
with normalization $\sum_r | J_r(\alpha)| = 1$,
\begin{eqnarray}
H_1(\alpha)&=&\sum^{2n-1}_{r=1}  J_r(\alpha) \sum^{4n}_{p=1}\vec{s}_{p}.\vec{s}_{p+r} \nonum\\ 
&+&J_{2n}(\alpha) \sum^{2n}_{p=1} \vec{s}_p.\vec{s}_{p+2n}  
\label{eq3}
\end{eqnarray}
in a chain of $N = 4n$ spins with periodic boundary conditions (PBC). \\

\begin{figure}[h]
\begin {center}
\hspace*{-0cm}{\includegraphics[width=8.5cm,height=2.50cm,angle=-0]{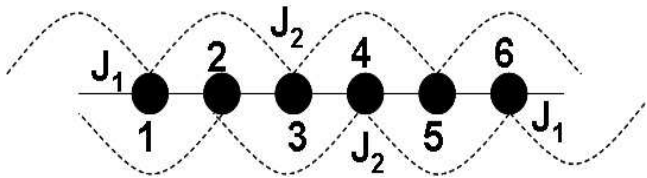}} \\
\caption{Schematic representation of a frustrated $s = 1/2$ spin chain with isotropic exchange $J_1 > 0$
between first neighbors and $g=J_2 > 0$ between second neighbors. }
\label{fig1}
\end {center}
\end{figure}
The $J_1-J_2$ model sketched in Fig. \ref{fig1} is the short-range $(1/\alpha = 0)$ limit of $H(g,\alpha)$ 
with $J_r = \delta_{1r}$. Using exact
diagonalization (ED) of finite systems, Sandvik \cite{r10} constructed an 
approximate quantum phase diagram of $H(g,\alpha)$ in the $(g,1/\alpha)$ plane; 
increasing $g$ leads to a first order transition for $\alpha < \alpha_c \approx 1.8$, 
a continuous transition for $\alpha > \alpha_c$. We also use ED and focus on systems
with large frustration $g$ where we identify and characterize a decoupled phase for any 
$\alpha$. We compare our phase 
diagram in the $(g,1/\alpha)$ plane with Sandvik's 
in Section III. \\

All models $H(g,\alpha)$ conserve total spin $S$ and have 
inversion symmetry $\sigma$ at sites. The ground state is always a singlet, $S = 0$.
Two excitations have special roles in the following: the magnetic gap $E_m(g,\alpha)$ 
to the lowest triplet state, $S = 1$, and the gap $E_{\sigma}(g,\alpha)$ 
to the lowest singlet with opposite inversion symmetry to the GS. In finite 
systems with fixed $\alpha$, the relation $E_{\sigma}(g,\alpha) = 0$ defines points $g$ 
at which the GS is doubly degenerate and states with broken inversion
symmetry can readily be constructed. The BOW phase of extended systems has 
$E_{\sigma}(g,\alpha) =0$ over some interval [$g$*,$g$**]. Since $E_m(g,\alpha)$ 
is known to open very slowly on entering a BOW phase \cite{r11,r6}, phase 
boundaries $g$* in finite systems have been inferred from level
crossing, the excited-state degeneracy $E_m$($g$*,$\alpha$) = $E_{\sigma}(g$*,$\alpha)$. 
For the $J_1-J_2$ model, Okamoto and Nomura \cite{r5} obtained $g$*=0.2411 
for the BOW phase, also called the dimer phase \cite{r7} or a 
valence bond solid (VBS)\cite{r10}. The excited-state degeneracy \cite{r12} 
leads in the $J_1-J_2$ model to $g$**= 2.02(3) for the boundary of the decoupled 
phase that is the focus of this paper. We find narrower BOW phases when $H_1(\alpha)$ 
has exchange beyond $J_1$ but no LRO. In Section II we solve exactly a model with 
uniform exchange that undergoes a first order quantum transition to a decoupled 
phase at $g_c = 1/4{\rm ln2}$. Models $H(g,\alpha)$ with variable $\alpha$ allow a
more complete characterization of systems with strong frustration. \\

We note that 
finite $g$** for the $J_1-J_2$ model disagrees with the field theory of White
and Affleck\cite{r7} or of Itoi and Qin \cite{r13}. Both start with small $g$ 
and infer finite $E_m$ for arbitrarily large $g$, albeit $E_m$ is exponentially 
small and has different $g$ dependence in the two treatments. At large frustration, 
it is natural to consider $H(g,\alpha)/g$ in Eq. \ref{eq1} as an HAF on
each sublattice. The $J_1-J_2$ model in Fig. \ref{fig1} then has 
nearest-neighbor $J_1 = 1/g$. Each spin $s_{2p}$ 
in sublattice A is coupled to two spins, $s_{2p+1}$ and $s_{2p-1}$, 
in sublattice B, and vice-versa. In the two-leg spin ladder, each spin in one 
leg (sublattice) is coupled to one spin in the other leg. Barnes et al. \cite{r14} 
conclude that arbitrarily small $J_{\perp} = 2/g$ at rungs opens a
finite gap in two-leg ladders, just as does any dimerization $J_1 = (1 \pm \delta)$ 
along the chain \cite{r4,r15}. Either $\delta$ or $J_{\perp}$ leads to two spins 
per unit cell and breaks inversion symmetry at sites. We contrast in Section III 
the magnetic gaps $E_m$ of the $J_1-J_2$ model and two-leg ladders.\\

In addition to the excitations $E_m(g,\alpha)$ and $E_{\sigma}(g,\alpha)$, 
we will focus on the GS expectation value of sublattice spin,
\begin{eqnarray}
\langle S^{2}_A \rangle = \langle S^{2}_B \rangle =-4n \sum^{n}_{p=1} \langle \vec{s}_1. \vec{s}_{2p} \rangle 
\label{eq4}
\end{eqnarray}
where we have used $S = 0$ and PBC. Correlation functions between spins in different
sublattices are governed by $H_1(\alpha)$ at $g = 0$ and become vanishingly small at large $g$.
Sublattice spin is an approximate or hidden symmetry since states with different $S_A$, $S_B$
are orthogonal. The GS is a linear combination of states centered on $\langle S^2_A \rangle^{1/2}$ 
that shift to smaller $S_A$ with increasing frustration.\\

The paper is organized as follows. The model solved in Section II has uniform
exchanges in $H_1$ and conserved $S_A$ and $S_B$. The GS has perfect LRO for $g< g_c(4n)$ 
that depends weakly on size and goes to $g_c = 1/4\rm{ln2}$ in the infinite chain. 
Increasing $g$ induces a first order transition to the decoupled phase that 
corresponds to noninteracting HAFs on sublattices. Accordingly, $E_{\sigma}$ and $E_m$ 
are directly related to HAF excitations. In Section III we present ED results for 
models $H(g,\alpha)$ that do not conserve $S_A$ or $S_B$. A first order transition 
to the decoupled phase is inferred for models with LRO at small frustration
based on almost identical $\langle S^2_A \rangle$, $E_{\sigma}$ and $E_m$ 
as in the uniform-exchange model. The BOW and decoupled phases of the $J_1-J_2$ 
model are related to degeneracies of ground states at $E_{\sigma} =0$ 
and of excited states at $E_{\sigma} = E_m$. The BOW/decoupled boundary of $H(g,\alpha)$ 
is estimated in the $(g,1/\alpha)$ plane. The magnetic gap $E_m$ in the decoupled phase 
is contrasted to $E_m$ of the two-leg ladder using the density matrix 
renormalization group (DMRG). In Section IV we briefly discuss the decoupled phase.

\section{Frustrated chain with uniform exchange}
In this Section, we solve a 1D model with $4n$ spins that conserves 
sublattice spins $S_A$ and $S_B$. Extensive results for finite and 
infinite linear HAFs are directly applicable in this case. The $H_1$ 
part, Eq. \ref{eq3}, is taken to have uniform AF exchange $J_r = 2/(4n-1)$  
between spins in opposite sublattices, as in the Lieb-Mattis model \cite{r16p}, and also uniform exchange $-J_r$ 
between all spins on the same sublattice. The $(4n-1)/2$ exchanges $|J_r|$ 
per site are normalized to unity. Uniform exchange makes it straightforward 
to express $H_1$ for $4n$ spins as
\begin{eqnarray}
H_1(4n)=\frac{S^2-2S_A^2-2S_B^2+3n}{(4n-1)}
\label{eq5}
\end{eqnarray}
The integer ranges are $0 \leq S \leq 2n$, $0 \leq  S_A$, $S_B \leq n$, and the 
index $\alpha$ has been omitted. The GS is evidently always in the $S = 0$ 
sector. In the absence of frustration, the GS is a linear combination of 
$2n + 1$ sublattice states with $S_A = S_B = n$ and $z$ components 
$M_B = -M_A$. The degeneracy of states with fixed $S$, $S_A$ and $S_B$ 
is lifted by $gH_2$ with frustrating $J_2 = g$. There are two $J_2$ 
contributions in Eq. \ref{eq1}, $g$ from $H_2$ and $-2/(4n - 1)$ from $H_1$. 
All eigenstates of the model are products of HAF eigenstates on sublattices. 
Increasing frustration generates energy shifts and numerous level crossings 
that can be followed explicitly.\\

We classify the HAF eigenvalues in Eq. \ref{eq2} as $E_i(S_A,2n) + E_j(S_B,2n)$ where 
$i$, $j$ = 0,1, 2, .. are eigenvalues with sublattice spins $S_A$ and $S_B$. 
The lowest energy $E_0(S,2n)$ in each 
sector is sufficient. The singlet GS for $4n$ spins in the singlet sector $S_A = S_B$ has energy
\begin{eqnarray}
E(0,S_A,g)=-\frac{4S_A(S_A+1)-3n}{4n-1}+2gE_0(S_A,2n). 
\label{eq6}
\end{eqnarray}
There is perfect ferromagnetic order with $S_A = n$ at small $g$, where 
$E_0(n,2n) = n/2$, and the GS transforms as $ \sigma=1$. The GS is doubly degenerate at $E(0,n,g) = E(0,n-1,g)$; 
the GS in the sector with $S_A =S_B < n$ is odd under inversion. Repeating the argument 
shows that the GS for $S_A = n-m$ is even, odd under inversion for even, odd $m$. Since inversion symmetry 
changes $n$ times between $S_A = n$ and $S_A = 0$, the GS for $S_A = 0$ transforms as $\sigma = (-1)^n$ 
and corresponds to the product  $|G\rangle |G \rangle$, the singlet GS of each sublattice.\\

We find $g_S(4n)$ at which the ordered state is degenerate with the GS in the sector $S = S_A < n$. 
The points are defined by $E(0,n,g) = E(0,S_A,g)$  
\begin{eqnarray}
g_S(4n)=\frac{n+1}{4n-1}\frac{(1-S_A(S_A+1)/n(n+1))}{(-\epsilon(S,2n)+1/4)}
\label{eq7}
\end{eqnarray}
where $\epsilon(S,2n) = E_0(S,2n)/2n$ is an energy per site and $S = 0, 1, 2, ..., n-1$. ED 
for $2n$-spin HAFs yields exact $g_S(4n)$ in Table 1 up to $4n \approx 60$. Increasing $g$ 
leads directly from $S_A = n$ to $S_A = 1$ at $g_1(4n)$ and then to $S_A = 0$ at $g_0(4n)$. 
Exact results in Eq. \ref{eq7} for the infinite chain place the first-order transition at 
\begin{eqnarray}
g_c=\frac{1}{4 \rm ln2}  .
\label{eq8}
\end{eqnarray}
The Bethe ansatz was used by Hulthen \cite{r1} to find the GS energy 
per site of the extended HAF; it has more recently been applied to finite 
systems of $N = 4n$ spins. Woynarovich and Eckle\cite{r16} found 
logarithmic corrections to the lowest 
energy per site in sector $S$. To leading order in $1/n$,  
\begin{eqnarray}
\epsilon(S,2n)-\epsilon(0,2n)=\frac{\pi^2S^2}{8n^2}\bigg(1-\frac{1}{2 {\rm ln}2n}\bigg).
\label{eq9}
\end{eqnarray}
The $n = 100$ entries in Table \ref{tb1} illustrate the convergence of 
$g_1(4n)$ and $g_0(4n)$. It follows from substituting Eq. \ref{eq9} into Eq. \ref{eq7} that 
the infinite chain also has $g_1 < g_0 < g_S$, $S \ge 2$.\\

The degeneracy $E(0,1,g)=E(0,0,g)$ between the GS in the singlet sector with $S_A = S_B  = 1$ 
and 0 occurs at
\begin{eqnarray}
g_{10}(4n)=\frac{4}{(4n-1)\epsilon_{ST}(2n)}.
\label{eq10} 
\end{eqnarray} 
The singlet-triplet gap, $\epsilon_{ST}(2n) = E_0(1,2n)- E_0(0,2n)$, 
appears frequently in the following. ED returns the $g_{10}$ entries in Table \ref{tb1}. 
To leading order for large systems, $\epsilon_{ST}$ is given by setting $S = 1$ and multiplying 
Eq. \ref{eq9} by $2n$, and then substituting in Eq. \ref{eq10}
\begin{eqnarray}
g_{10}(4n>>1)=\frac{4}{\pi^2}\bigg(\frac{4n}{4n-1}\bigg)\frac{1}{(1-1/2 {\rm ln}2n)}.
\label{eq11} 
\end{eqnarray} 
The absolute GS for $g > g_{10}(4n)$ is $E(0,0,g)$ for chains of any length. 
The eigenstate $|G \rangle |G \rangle$ has $S_A = S_B = 0$, vanishing spin correlations 
between sublattices in Eq. \ref{eq4} and hence no possibility of a BOW phase. 
The system of 400 spins in Table \ref{tb1} shows slow convergence to $g_{10} = 4/\pi^{2}$. \\
\begin{table}
\begin{center}
\caption {Ground state degeneracies of the frustrated chain with 
$N=4n$ spins and uniform exchange. The GS with $S_A = n$ and $S_A = 0$ or 
1 are degenerate at $g_0(4n)$ or $g_1(4n)$, respectively; the GS at 
$S_A = 0$ and 1 are degenerate at $g_{10}(4n)$.}
\begin{tabular}{cccc} \hline
$N=4n$&  $g_0$ & $g_1$ & $g_{10}$ \\\hline
~~~~~~~~~ 16 $~~~~~~~$       &0.47189 $~~~~~$      &0.46789 ~~~~~~&0.51020\\
~~~~~~~~~ 20 $~~~~~~~$       &0.45013 $~~~~~$      &0.44710 ~~~~~~&0.49735\\
~~~~~~~~~ 24 $~~~~~~~$       &0.43544 $~~~~~$      &0.43308 ~~~~~~&0.48873\\
~~~~~~~~~ 28 $~~~~~~~$       &0.42486 $~~~~~$      &0.42299 ~~~~~~&0.48341\\
~~~~~~~~~ 32 $~~~~~~~$       &0.41689 $~~~~~$      &0.41539 ~~~~~~&0.47756\\
~~~~~~~~~ 36 $~~~~~~~$       &0.41063 $~~~~~$      &0.40944 ~~~~~~&0.47382\\
~~~~~~~~~ 40 $~~~~~~~$       &0.40570 $~~~~~$      &0.40467 ~~~~~~&0.47058\\
~~~~~~~~~ 400 $~~~~~~~$      &0.36518 $~~~~~$      &0.36510 ~~~~~~&0.44864\\
~~~~~~~~~$\infty$$~~~~~~$  &0.36067   $~~~~~$      &0.36067 ~~~~~~&0.40528\\\hline 
\end{tabular}
\label{tb1}
\end{center}
\end{table}

The interval between $g_1(4n)$ and $g_{10}(4n)$ is not relevant in the 
context of the infinite chain, in which $\langle S^2_A  \rangle = n(n + 1)$
 for arbitrarily large $n$ drops to $S^2_A = 2$ at $g_c = 1/4{\rm ln2}$ 
and vanishes for $g \ge g_{10} = 4/\pi^2$. The discontinuity at $g_c$ marks a first 
order transition to the decoupled phase. Spin correlations between the sublattices 
vanish rigorously for $g \ge 4/\pi^2$. The degeneracy at $g_{10}(4n)$ involves states 
of opposite inversion symmetry, as does degeneracy at $g_1(4n)$ for even $n$.  For odd $n$, 
however, the GS in the sectors $S_A = n$ and 1 are both even under inversion. They are 
mixed and lead to an avoided crossing in finite models that do not conserve $S_A$.\\ 

Next we find $E_m(g)$ and $E_\sigma(g)$, the excitation energy to the 
lowest triplet and the lowest singlet with reversed inversion symmetry. It 
follows from Eq. \ref{eq5} that the lowest triplet for $g < g_{10}(4n) $ 
is obtained by changing $S$ from 0 to 1 without changing $S_A$ or  $S_B$; hence $E_m(g,4n) = 2/(4n-1)$ 
is constant, independent of $g$ up to $g_{10}(4n)$. When the GS energy is $E(0,0,g)$, 
the lowest triplet has $S = 1$, $S_A + S_B = 1$. It is doubly degenerate, $|G \rangle |T \rangle$ 
or $|T \rangle |G \rangle $ in obvious notation, with a triplet on either sublattice. The magnetic gap is
\begin{eqnarray}
E_m(g,4n)=g\epsilon_{ST}(2n)-2/(4n-1)  ~~~~~ g\geq g_{10}(4n). 
\label{eq12} 
\end{eqnarray} 
The second term is the contribution from Eq. \ref{eq5}.\\
 
The lowest singlet excitation for $g > g_1(4n)$ is
\begin{eqnarray}
E_{\sigma}(g,4n)&=&|E_0(0,1,g)-E_0(0,0,g)| \nonum \\
&=&|2g \epsilon_{ST}(2n)-\frac{8}{(4n-1)}|.
\label{eq13} 
\end{eqnarray} 
The excitations $E_m(g)$ in Eq. \ref{eq12} and $E_{\sigma}(g)$ 
in Eq. \ref{eq13} are equal at $g = 3g_{10}(4n)/2$. It is instructive 
to rewrite the excitations using Eq. \ref{eq10}
\begin{eqnarray}
E_{m}(g,4n)&=&\frac{4}{4n-1}\bigg(\frac{g}{g_{10}(4n)}-\frac{1}{2}\bigg)~~~ g \geq g_{10}(4n) \nonum \\
E_{\sigma}(g,4n)&=&\frac{8}{4n-1}\bigg|\frac{g}{g_{10}(4n)}-1\bigg| ~~ g \geq g_{1}(4n). 
\label{eq14} 
\end{eqnarray} 
The V-shaped dependence of $E_{\sigma}(g,4n)$ on either side of 
$g_{10}$ is evident, as are the related slopes with increasing $g$. 
These features are used in Section III to interpret $E_m$ and $E_{\sigma}$ 
in models that do not conserve $S_A$. It is still convenient 
to refer to products of $|G \rangle$ or $|T \rangle $. Although no 
longer exact $1/g = 0$ eigenstates, the actual eigenstates can be 
expanded in terms of sublattice eigenstates.\\

We conclude the discussion of the uniform exchange model by noting 
that the infinite chain has a first order quantum 
transition at modest frustration $g_c = 1/4 \rm ln2$. The GS has 
perfect LRO up to $g_c$. The decoupled phase has $S_A = S_B = 1$ 
in the interval $g_c < g < g_{10} = 4/\pi^2$ and $S_A = S_B = 0$ 
for $g > g_{10}$, when all spin correlation functions 
in Eq. \ref{eq4} are zero. The infinite chain has $E_m(g) = 0$ for 
all $g \ge 0$.
\section{Frustrated chains with variable range exchange} 
We present ED results for models $H(g,\alpha)$ that do not conserve sublattice spin. 
The GS is degenerate under inversion at frustration $g_j$ where $E_{\sigma}(g_j,4n) = 0$. 
It is convenient to retain the labeling $g_S(4n)$ and $g_{10}(4n)$ in Eqs. \ref{eq7} and \ref{eq10} 
used for the model with uniform exchange. We start with a model with LRO at $g = 0$ 
and a first order transition to the decoupled phase that closely resembles the 
uniform model. Next we consider the $J_1-J_2$ model without LRO at $g = 0$ and 
continuous transitions with increasing $g$ from spin liquid to BOW to decoupled phase. 
We then consider intermediate $\alpha$ to construct an approximate the GS phase 
diagram in the (g, $1/\alpha$) plane.\\
\subsection{Model with LRO} 
The Hamiltonian $H(g,\alpha)$ in Eq. \ref{eq1} has nonfrustrating exchanges in Eq. \ref{eq3}. 
It differs from the model studied by Sandvik only in the $J_{2n}$ terms, which are double counted in Eq. \ref{eq1} 
of ref \cite{r10}. Since $J_{2n}$ contributions decrease 
with size $n$ or increasing exponent $\alpha$, numerical difference due to double counting are 
limited to small $n$ and $\alpha$. The normalization condition $\sum_r|J_r| = 1$ leads to
\begin{eqnarray}
J_{r \neq 2 }(\alpha) &=&\frac{(-1)^{r-1}}{r^{\alpha}}\bigg(1+\frac{1}{2}(\frac{1}{2n})^{\alpha} \nonum  \\
&+& \sum^{2n-1}_{s=3} \frac{1}{s^{\alpha}}\bigg)^{-1}. 
\label{eq15} 
\end{eqnarray} 
\begin{figure}[h]
\begin {center}
\hspace*{-0cm}{\includegraphics[width=7.5cm,height=8.5cm,angle=-90]{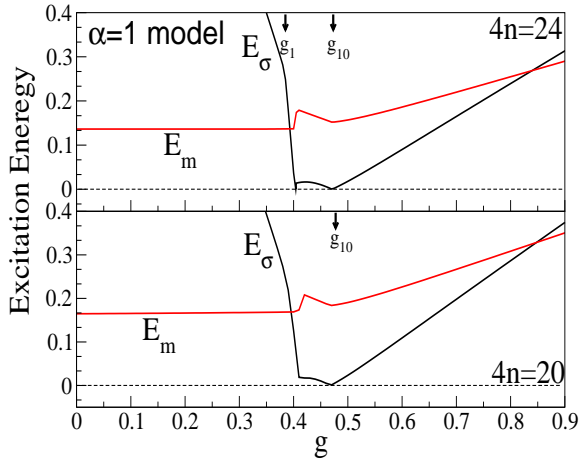}} \\
\caption{ Excitation energies $E_m$ and $E_{\sigma}$ as functions of frustration $g$ 
in models with exchanges with $\alpha = 1$ in Eq. \ref{eq15} and $4n = 24$ 
or 20 spins; $E_m$ is to the lowest triplet, $E_{\sigma}$ to the 
lowest singlet with reversed inversion symmetry. The ground state is doubly 
degenerate at $g_{10}$ where $E_{\sigma}(g_{10}) = 0$ and at $E_{\sigma}(g_1)$ 
for 24 spins. For $g \geq g_{10}$, the triplet is doubly degenerate 
and the excited singlet is $^1|T\rangle |T\rangle$, a triplet on each sublattice.}
\label{fig2}
\end {center}
\end{figure}
Since frustrating $J_2 = g$ is entirely in $gH_2$, the model with 
$\alpha = 0$ and finite $n$ in Eq. \ref{eq15} is slightly different from the 
model with uniform exchange.\\

The following results are for $\alpha = 1$, 
a model with \cite{r8} LRO at $g = 0$.
Excitation energies $E_m(g)$ and $E_{\sigma}(g)$ are shown in Fig. \ref{fig2} as 
a function of $g$ for $4n = 24$ sites (top panel) and 20 sites (bottom panel). As 
anticipated for LRO systems with a first order transition, we have $E_{\sigma}(g) = 0$ 
at two points $g_1$ and $g_{10}$ when $n$ is even and one point $g_{10}$ when $n$ 
is odd. Inversion symmetry at sites reverses twice for 24 sites, once for 20 sites. 
The first order transition is at $g_1 = 0.404$ for 24 spins and  the avoided crossing is 
at $g = 0.411$ for 20 spins. The degeneracy at $g_{10}$ corresponds to changing from $S_A = 1$ 
to 0 in the model with uniform exchange. The GS in the $\sigma = 1$ and -1 sectors cross at $g_{10}$ 
for $4n = 20$, in contrast to the avoided crossing in Fig. \ref{fig2} of ref. \cite{r10} for the 
$\alpha = 1$ model with \cite{r17} $4n = 28$. \\

The separation $g_{10} - g_1 = 0.065$ for the $\alpha = 1$ 
model is comparable to $g_{10}(24) - g_1(24) = 0.056$ in Table \ref{tb1} for uniform exchange. 
In accord with Eq. \ref{eq12} for uniform exchange, $E_m(g)$ is almost constant 
up to $g_1$; it is doubly degenerate for $g \ge g_{10}$ and linear with increasing $g$. 
The slopes $dE_m/dg$ are within $1\%$ of $\epsilon_{ST}(12) = 0.356$ for 24 sites and 
$\epsilon_{ST}(10) = 0.423$ for 20 sites. Likewise for the lowest singlet excitation, 
$^1|T \rangle |T \rangle$, the slope of $E_{\sigma}(g)$ between $g_{10}(4n)$ and 
$2g_{10}(4n)$ is slightly larger than $2\epsilon_{ST}(2n)$ in either case, 
$dE_{\sigma}/dg = 0.741$ for 24 sites and 0.884  for 20 sites. \\ 

\begin{figure}[h]
\begin {center}
\hspace*{-0cm}{\includegraphics[width=7.5cm,height=8.5cm,angle=-90]{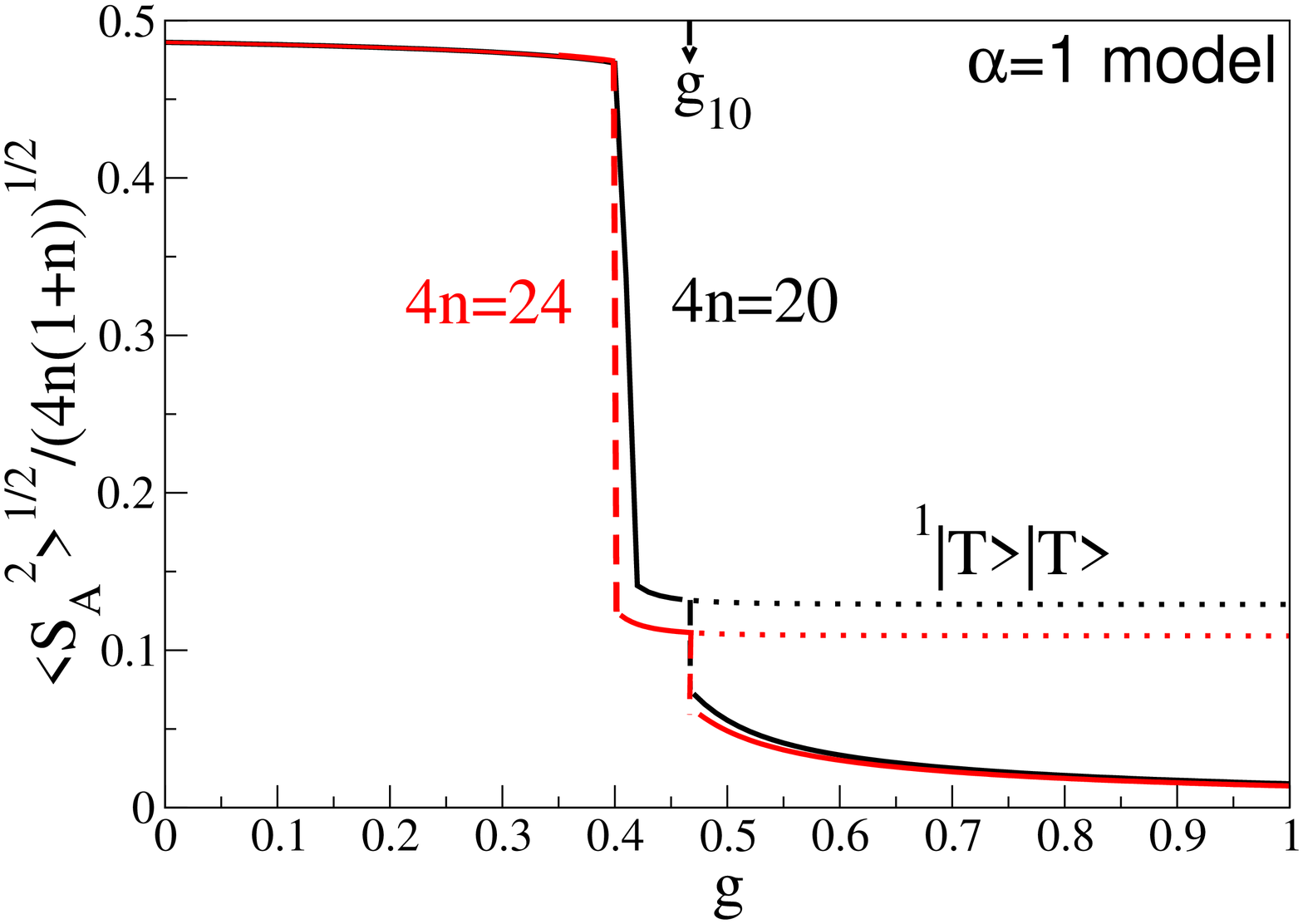}} \\
\caption{Solid lines: Ground state expectation value $s_A(g,4n)$ in Eq. \ref{eq16} 
with increasing frustration $g$ in models with 20 and 24 spins and exchanges $\alpha = 1$ 
in Eq. \ref{eq15}. Dotted lines for $g \ge g_{10}$ are $s_A(g,4n)$ for 
the excited singlet $^1|T\rangle |T \rangle$.}
\label{fig3}
\end {center}
\end{figure}

We define $s_A(g,4n) \le 1/2$ to quantify sublattice spin as the GS expectation value
\begin{eqnarray}
s_A(g,4n)=\langle S^2_A \rangle^{1/2}/(4n(n+1))^{1/2}. 
\label{eq16} 
\end{eqnarray} 
$s_A(g,4n)$ decreases with frustration and is double-valued at 
$E_{\sigma}(g_j) = 0$. Figure \ref{fig3} shows $s_A(g,4n)$ with 
increasing $g$ for 20 and 24 sites. Dotted lines for 
$g > g_{10}(4n)$ refer to the excited state $^1|T \rangle |T \rangle$. 
The $s_A(g_{10},4n)$ values in Table \ref{tb2} at modest 
frustration $g \approx 0.5$ are already close to  $1/[2n(n+1)]^{1/2}$, 
the exact result at $1/g = 0$. The $\sigma = 1$ eigenstates near 
the transition contain small admixtures of sublattice spin. 
The $\alpha= 1$ model requires matrix elements for all $J_r$ in 
Eq. \ref{eq3}, $r = 1,2, ... 2n$ that make it tedious to evaluate $\langle S^2_A \rangle$. 
The results in Table \ref{tb2} go to 24 spins for the $\alpha= 1$ 
\begin{table}
\begin{center}
\caption {Ground and excited state values of sublattice spin, 
$s_A(g_{10},4n)$ in Eq. \ref{eq16}, for the $\alpha=1$ and $J_1-J_2$ models.}
\begin{tabular}{cccccc} \hline
Model        &  ~~~~$\alpha=1 $   ~~~~ & $J_1-J_2    $ & $\alpha=1$ & $J_1-J_2$  & $1/g=0$  \\
4n/state     &  ~~~~$|GS \rangle $~~~~ & $|GS \rangle$ & $^1|T \rangle |T\rangle$  
& $^1|T \rangle |T\rangle$ & $^1|T \rangle |T\rangle$  \\\hline
16 & 0.092 &  0.106  & 0.161  & 0.150  & 0.158 \\
20 & 0.081 &  0.088  & 0.132  & 0.123  & 0.130 \\
24 & 0.062 &  0.075  & 0.110  & 0.104  & 0.109 \\
28 & –     &  0.067  &  –     & 0.093  & 0.094 \\\hline
\end{tabular}
\label{tb2}
\end{center}
\end{table}
model and to 28 spins for the $J_1-J_2$ model. Somewhat larger 
systems are accessible in principle with current computational resources.\\

The $\alpha= 1$ model has almost perfect LRO at $g = 0$, $s_A(0,4n) = 0.486$, 
in agreement with larger systems studied in ref. \cite{r8} using multiple methods. 
As anticipated in Section II, $s_A(g,4n)$ is discontinuous when $E_{\sigma}(g) = 0$ 
and changes rapidly but continuously for 20 spins at the avoided crossing. 
The size dependence of $s_A(g_{10},4n)$ shown in Fig. \ref{fig3} is consistent 
with a discontinuity at $g_1$ in the infinite system and $s_A(g_{10}) \approx 0$ for $g > g_{10}$. 
Both results can be understood in terms of a GS with LRO at small $g$ 
and a first order transition to the decoupled phase. \\ 

\begin{figure}[h]
\begin {center}
\hspace*{-0cm}{\includegraphics[width=7.5cm,height=8.5cm,angle=-90]{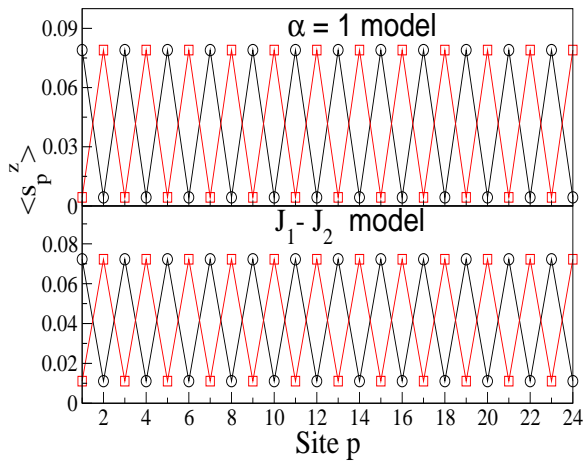}} \\
\caption{Spin density $\rho_p$ at site $p$ in the lowest triplet of 
24 spin chains at frustration $g_{10}$, the largest $g$ 
at which the ground state is degenerate. The upper panel has 
exchanges $\alpha = 1$ in Eq. 15 (top); the lower panel is the $J_1-J_2$ model.}
\label{fig4}
\end {center}
\end{figure}

All properties of frustrated spin chains of $4n$ spins are given at $1/g = 0$ by exact HAF 
eigenstates for $2n$ spins. The triplet $|T \rangle$ with $S^z = 1$ and energy $E_0(1,2n)$ 
has equal spin density $\rho_p = s^z_p = 1/2n$ at all sites. The spin 
density at site $p$ of the degenerate triplets $|G \rangle|T \rangle$  
and 
$|T \rangle |G \rangle$ at $1/g = 0$ is
\begin{eqnarray}
\rho_p(4n)=\frac{1}{4n}(1\pm(-1)^p).  
\label{eq17} 
\end{eqnarray} 
The plus sign corresponds to a triplet on the sublattice of even-numbered sites and $\rho_p = 0$ 
for odd $p$; the minus sign has the triplet on the odd-numbered sublattice. 
The $\alpha = 1$ spin densities at $g_{10}$ are shown in the top panel of Fig. \ref{fig4} 
for 24 spins. As expected, $\rho_p$ is constant on each sublattice, slightly 
less than $1/2n$ on one and slightly positive on the other. The deviations 
from Eq. \ref{eq17} are less $0.1\%$ at $g = 1$, less than $0.01\%$ at $g = 2$. 
The decoupled phase found rigorously for uniform exchange readily 
accounts for the $\alpha = 1$ model.\\  
\subsection{$J_1-J_2$ model}
The linear HAF with $J_1$ between nearest neighbors and frustration $g = J_2/J_1$ 
is a prototypical model with a BOW phase \cite{r5,r6,r7}. The model evolves 
from an HAF at $g = 0$ to two HAFs at $1/g = 0$. Okamoto and Nomura \cite{r5} 
placed the continuous transition to the BOW phase at $g$*=0.2411 by extrapolation 
of the excited state degeneracy $E_{\sigma}(g,4n) = E_m(g,4n)$ up 
to $4n = 24$ spins. The GS is nondegenerate for $g < g$*. There is no 
LRO at $g = 0$ but there are long-range spin correlation functions that go as \cite{r8} 
\begin{eqnarray}
\langle \vec{s}_0.\vec{s}_p \rangle = \frac{(-1)^p \sqrt{{\rm ln}p}}{p}. 
\label{eq18} 
\end{eqnarray} 
The exact values \cite{r18} of $\langle \vec{s}_0.\vec{s}_1 \rangle$ and 
$\langle \vec{s}_0.\vec{s}_3 \rangle$ are $1/4- \rm ln2 = -0.443147..$ and $-0.15074...$, 
respectively, and all correlations between spins in opposite sublattices 
are negative. It follows that $\langle S^2_A \rangle$ for $4n$ spins in Eq. \ref{eq4} is of 
order $4n {\rm ln}4n$ at $g=0$. \\

\begin{figure}[h]
\begin {center}
\hspace*{-0cm}{\includegraphics[width=7.5cm,height=8.5cm,angle=-90]{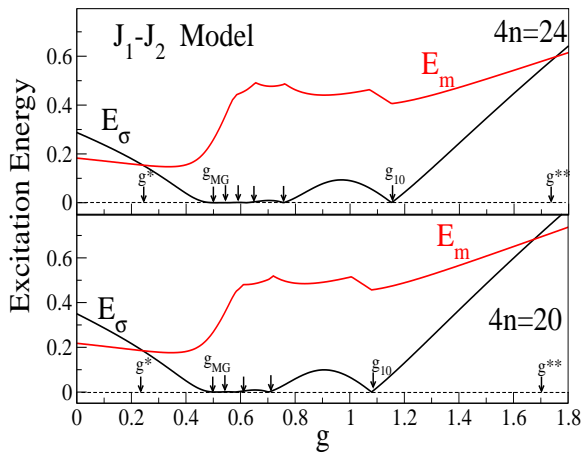}} \\
\caption{Excitation energies $E_m$ and $E_{\sigma}$ with increasing frustration $g$ in $J_1-J_2$ 
models of $4n = 24$ and 20 spins; $E_m$ is to the lowest singlet and $E_{\sigma}$ 
is to the lowest singlet with reversed inversion symmetry. Arrows indicate $g_j$ 
where the ground state is doubly degenerate, $E_{\sigma}(g_j) = 0$;
 points $g$* and $g$** mark $E_m = E_{\sigma}$.}.
\label{fig5}
\end {center}
\end{figure}

The $J_1-J_2$ model has multiple GS degeneracies \cite{r12} $E_{\sigma}(g,4n) = 0$ 
with increasing $g$. The first one at $g_{MG} = 1/2$ is the Majumdar-Ghosh 
point \cite{r19} where the exact GS is known for any even number of spins. 
Spin correlation functions are now limited to nearest neighbors, and the 
infinite chain has $\langle S^2_A \rangle = 3n/2$, or 3/8 per site. In contrast to the transition 
from $S_A = n$ to $S_A = 1$ in models with LRO at $g = 0$, the symmetry 
changes at $g_j(4n), j = n, n-1,... $, occur 
sequentially with increasing $g$ up to 28 spins, the largest $4n$ system we 
solved. The upper panel of Fig. \ref{fig5} shows $E_{\sigma}(g,24)$ and $E_m(g,24)$ as a function 
of frustration, with six arrows at $g_j(24)$. The magnitude of $E_{\sigma}(g,4n)$ is 
remarkably small between $g \approx 0.45$ and $\approx 1.2$. A finer energy scale is 
needed to see $E_{\sigma}$ in the BOW phase. The lower panel shows $E_{\sigma}$ 
and $E_m$ for $4n = 20$ spins, with five arrows at $g_j(20)$. The behavior at large frustration 
is qualitatively similar to the $\alpha = 1$ model in Fig. \ref{fig2}. The triplet is 
doubly degenerate for $g > g_{10}$. The slopes $dE_m/dg$ and $dE_{\sigma}/dg$ are 
within $5\%$ and $8\%$ of $\epsilon_{ST}(2n)$ and $2\epsilon_{ST}(2n)$, 
respectively, between $g_{10}(4n)$ and $g = 2$. \\

We find that the relation $E_{\sigma}(g,\alpha) = 0$ is not limited to 
finite $J_1-J_2$ models. On the contrary, Sandvik \cite{r10} states that the 
degeneracy is not exact except in the $J_1-J_2$ model at the special point 
$g_{MG} = 1/2$. The discrepancy is not due to ED but to motivation. ED 
is necessarily performed at fixed values of $g$ that are typically on a grid. 
The GS symmetry changes in calculations that keep track of inversion at sites. 
It is then natural to search for the exact $(g,\alpha)$ at which the GS is 
degenerate by choosing $g$ more precisely. The lowest two 
singlets are both even under inversion at the center of bonds, which suggests 
an avoided crossing and less reason for refining $g$ in search of exact 
degeneracy. The symmetry operators of course commute with $H(g,\alpha)$ 
but not with each other. Hence inversion symmetry at sites or at bonds 
leads to different linear combinations of degenerate eigenstates. \\

Degenerate GS at $g_{MG}$ are products of singlet-paired spins on successive sites, 
either $|K_1 \rangle = (1,2)(3,4)...(2n-1,2n)$ or $|K2 \rangle = (2,3)(4,5)...(2n,1)$. 
Such states are the familiar Kekul\'e diagrams of organic chemistry 
or the degenerate GS of polyacetylene in the Su-Schrieffer-Heeger (SSH) 
model \cite{r20}. The gap $E_m$ that opens at $g$* 
is rigorously known \cite{r20p} to be finite in the infinite chain at $g_{GM}$. 
It is already large{\cite{r12} at $g_n(4n) = g_{MG}$ and remains large up 
to $g_{10}(4n)$, clearly exceeding finite-size effects in Fig. \ref{fig5}. The elementary excitations of 
the BOW phase are topological spin solitons \cite{r20p} or domain walls 
generated by spin correlations that closely resemble \cite{r12} SSH 
solitons generated by electron-phonon coupling. The BOW phase extends 
beyond $g_{10}(4n)$ in Fig. \ref{fig5}, the GS degeneracy at the largest frustration. 
We suppose that the BOW phase terminates at $g$**$\approx 2.02(3)$ at the excited-state degeneracy 
\cite{r12} $E_\sigma = E_m$. As discussed below, finite-size effects are larger at $g$** than at $g$*.\\
\begin{figure}[h]
\begin {center}
\hspace*{-0cm}{\includegraphics[width=7.5cm,height=8.5cm,angle=-90]{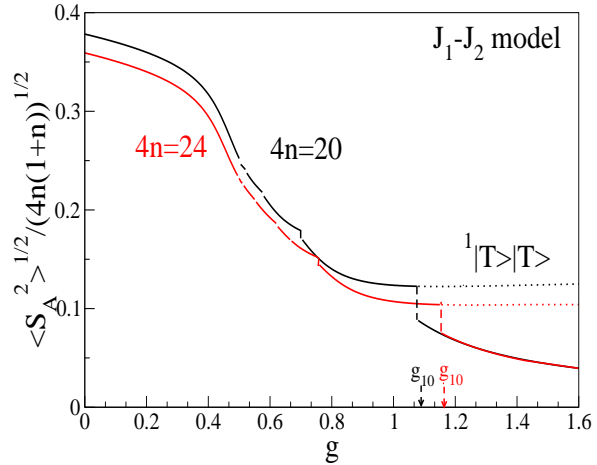}} \\
\caption{Solid lines: Ground state expectation value $s_A(g,4n)$ in Eq. \ref{eq16} 
with increasing frustration $g$ in $J_1-J_2$ models with 20 and 24 spins. 
Dotted lines for $g \geq g_{10}$ are $s_A(g,4n)$ for the excited singlet $^1|T \rangle |T \rangle$.}
\label{fig6}
\end {center}
\end{figure}
\begin{figure}[h]
\begin {center}
\hspace*{-0cm}{\includegraphics[width=7.5cm,height=8.5cm,angle=-90]{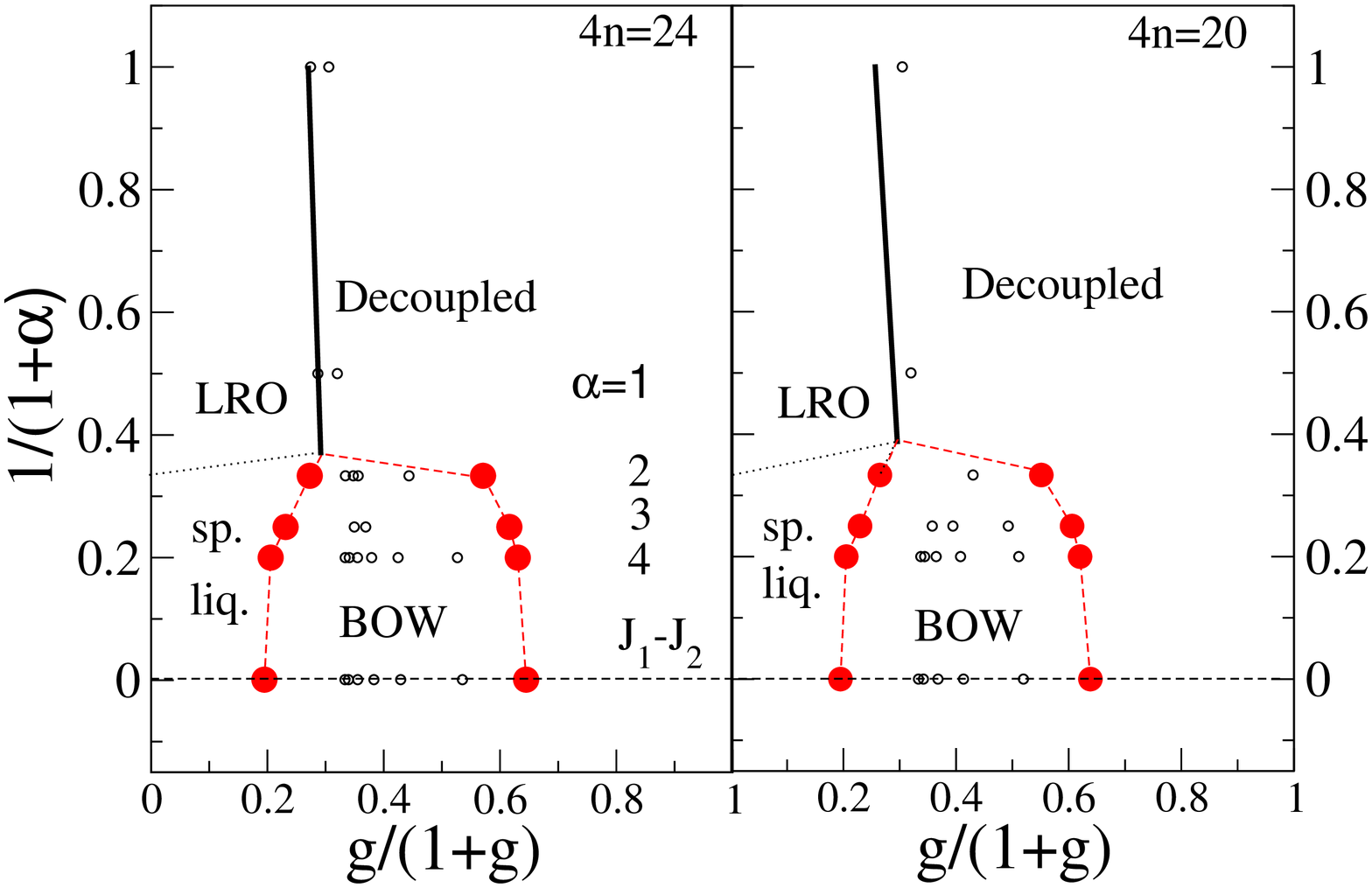}} \\
\caption{Quantum phase diagram of $H(g,\alpha)$ in Eq. \ref{eq1} in the $1/\alpha$, $g$ plane for chains 
of 20 and 24 spins; $g > 0$ is frustration and $\alpha$ in Eq. \ref{eq15} 
specifies the nonfrustrating exchanges. Open points and solid points indicate 
ground-state degeneracy $E_{\sigma} = 0$ and excited state degeneracy, $E_m = E_{\sigma}$, 
respectively. Dashed lines are approximate boundaries of the BOW phase. Solid 
lines are approximate boundaries of the decoupled phase in models with LRO at $g = 0$; 
the dotted line separates models with long-range fluctuations and order at small frustration \cite{r10}.} 
\label{fig7}
\end {center}
\end{figure}

The evolution of $s_A(g,4n)$ with increasing frustration is shown in Fig. \ref{fig6} for 20 
and 24 spins; $s_A(g,4n)$ for the excited state $^1|T \rangle |T \rangle$  is 
the dotted line for $g > g_{10}(4n)$. As expected, $s_A(g,4n)$ decreases with 
increasing $g$ and is discontinuous when the GS is degenerate. Table \ref{tb2} 
lists both values of $s_A(g_{10},4n)$. The excited state at $g_{10}$ is already within $5\%$ 
of the $1/g = 0$ limit. The $J_1-J_2$ model has stronger but still modest mixing 
of $1/g = 0$ states at $g_{10}$ than the $\alpha = 1$ model. The $J_1-J_2$ spin densities 
of the degenerate triplets at $g_{10}$ in Fig. \ref{fig4} again follow Eq. \ref{eq17} with $\rho_p$ 
slightly less than $1/2n$ on one sublattice and slightly positive on the other. 
The spin densities at $g = 2$ are $\rho_p = 0.082$ and $0.002$. At $g_{MG} = 1/2$, 
the overlap of the VB diagrams for 4n spins is $\langle K1| K2\rangle = 2^{-(2n-1)}$. 
Aside from finite-size effects that are already less than $1\%$ at $4n = 20$, we obtain
\begin{eqnarray}
s_A(1/2,4n) = {\bigg(\frac{3}{8(n+1)}\bigg)}^{1/2}. 
\label{eq19} 
\end{eqnarray} 
The infinite chain has continuous $s_A(g)$ over the entire range $g \ge 0$.
\subsection{Quantum phase diagram}
Models with $\alpha =$ 2, 3 and 4 have nonfrustrating exchange intermediate 
between $\alpha = 1$ and $1/\alpha = 0$. We find the GS degeneracies $g_m(4n)$ 
where $E_\sigma = 0$ and the excited state degeneracies $g$* and $g$** where $E_{\sigma} = E_m$ 
that delimit the BOW phase.These points are used to construct 
an approximate quantum phase diagram in the $(g,1/\alpha)$ plane. \\

Fig. \ref{fig7} shows the phase diagram of $H(g,\alpha)$ for $\alpha = 0$ to 
4 and the $J_1-J_2$ model $(1/\alpha = 0)$ over the entire range 
of frustration $g \ge 0$. The diagrams for 24 and 20 spins illustrate 
the modest size dependence and differences between $4n$ with even and odd $n$. 
Open points indicate GS degeneracy $g_m$; closed points are $g$* and $g$**; 
solid lines mark first order transitions at $g_1$ for 24 spins and the 
avoided crossing for 20 spins; dashed lines are the boundary between the BOW and decoupled 
phases at large $g$ and between the spin liquid and BOW phase at small $g$. 
The BOW phase terminates at a multicritical point, and the dotted line separates 
spin liquids from models with LRO. The BOW phase for $\alpha = 4$ closely resembles 
the $J_1-J_2$ model, as might be expected since largest change is small, $J_3/J_1 = 1/16$. 
The width of the BOW phase, $g$**-$g$*, narrows for $\alpha = 3$ or 2 and $E_{\sigma}= 0$ 
is satisfied at fewer than $n$ points. The $\alpha = 1$ model has a first order transition and no BOW phase.\\ 

Sandvik \cite{r10} used additional values of $\alpha$ to estimate the multicritical point 
as $\alpha_c \approx 1.8$, $g_c \approx 0.41$. We have not varied $\alpha$ and took $\alpha_c = 1.8$ 
in Fig. \ref{fig7}. It will be challenging to be more accurate as 
long as ED is limited to about 30 spins. Our phase diagram in Fig. 7 is quite 
similar at small frustration to Fig. \ref{fig1} of ref. \cite{r10} 
with VBS instead of BOW, AFM instead of LRO and QLRD($\pi$) instead of spin liquid. 
So far there is no consensus for naming phases. There are clear differences at large $g$, 
however. The VBS or BOW phase in Fig. 1 of ref. \cite{r10} does 
not terminate at either large $g$ or at small $\alpha$ and the line at large $g$ 
between VBS and VBS+QLRD($\pi/2$) separates phases with different $\alpha$. 
A BOW phase is rigorously excluded at $ \alpha = 0$, the analytical model with 
a first order transition to the decoupled phase.\\

We have given reasons for extending the decoupled phase
in Fig. \ref{fig7} to first order LRO/decoupled transitions and
to continuous BOW/decoupled transitions without any
distinction for different $\alpha$. The sublattices of 
$H(g, \alpha )$ have weak interactions at large
$g$. The coupling is weak even in the $J_1 - J_2$ 
 model and $s_A (g, 4n )$ becomes arbitrarily small at large $g$. 
By continuity, we expect the same phase to be reached at
$1/g << 1$ for any choice of $H(g, \alpha)$. The decoupled 
phase starts at the first order transition in systems with LRO
at $g  = 0$ and at $g$** in systems with a BOW phase 
and continuous transitions.\\
\subsection{Magnetic gap}

Density matrix renormalization group (DMRG) has been extensively applied to 1D spin systems 
\cite{r7,r21,r22,r23,r24}. DMRG with open boundary conditions (OBC) breaks inversion symmetry 
at sites at the outset: an even number of spins is required for a singlet GS and 
even chains have inversion symmetry at the center of the central bond. As shown 
explicitly for a half-filled band of free electrons \cite{r23}, OBC generates $1/N$ corrections 
to the bond order of the central bond. 
OBC strongly breaks the GS degeneracies of the $J_1-J_2$ model at $g_j(4n)$. At these points, 
the lowest singlet excitation for 16, 20 or 24 spins is slightly  {\it higher} than $E_m$, thus reversing 
the order of excitations in addition to lifting the degeneracy.\\ 

It is convenient to study $H(g,\alpha)/g$ in Eq. \ref{eq1} for 
$1/g \ll 1$. Then $H_2$ has decoupled HAFs on the sublattices for models with any exponent 
$\alpha$. In particular, the frustrated $J_1-J_2$ model has $N$ exchanges $J_1 = 1/g$ between spins at adjacent 
sites $p$, $p + 1$ in Fig. \ref{fig1} while the non-frustrated two-leg ladder has $N/2$ 
exchanges $J_{\perp} = 2/g$ between sites $2p - 1$, $2p$. The $J_1-J_2$ model has one spin 
per unit cell and inversion symmetry at both sites and centers of bonds. The two-leg ladder 
with $J_{\perp} > 0$ has two spins per unit cell and inversion only at bond centers. \\

DMRG results for $E_m$ are compared in Fig. \ref{fig8} for $J_1-J_2$ models and two-leg ladders, 
and are seen to be qualitatively different. The ladder has large $E_m = 0.49$ at $g = 2$ that 
decreases to $E_m = 0.20$ at $g = 4$. Finite $E_m$ is expected \cite{r14} for 
finite $J_{\perp}$, nearly linear in small $J_{\perp}$, just as for finite 
dimerization \cite{r4,r15} $\delta$ in chains with alternating $J_1 = (1 \pm \delta)$ and 
$J_2 = 0$. The $J_1-J_2$ model returns $E_m \approx 0.03$ at $g = 2$, 
close to the BOW/decoupled boundary $g$**, and $E_m < 0.01$ at 
$g = 3$ or 4, the limit \cite{r23} of accuracy for DMRG with four spins added per step. 
Large $g$ in Fig. \ref{fig8} brings out the contrasting behavior of $E_m$. While an 
exponentially small $E_m$ cannot be ruled out in the $J_1-J_2$ model, DMRG is consistent 
with $E_m \approx 0$ in the decoupled phase and suggests that $g$** is 
slightly larger than 2.0,
\begin{figure}[h]
\begin {center}
\hspace*{-0cm}{\includegraphics[width=7.5cm,height=8.5cm,angle=-90]{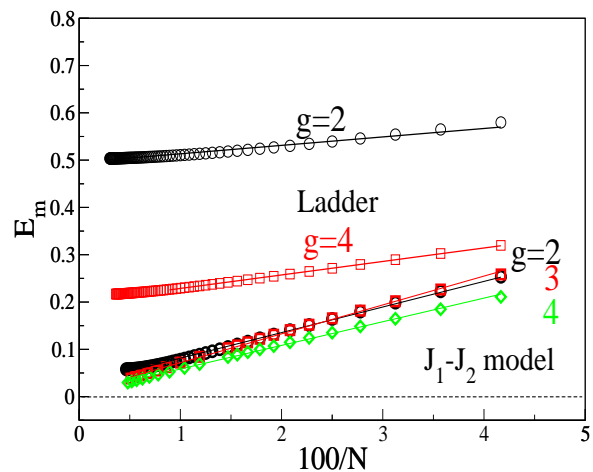}} \\
\caption{DMRG results the magnetic gap $E_m$ of $N = 4n$ spins for $J_1-J_2$ models with $J_1 = 1/g$, $J_2 = 1$ 
in Eq. \ref{eq1} and for two-leg ladders with $ J_{\perp} = 2/g$ at every rung.}
\label{fig8}
\end {center}
\end{figure}
the value estimated from extrapolation of $E_m = E_{\sigma}$.\\

The relation $E_{\sigma} = E_m$ for the BOW/decoupled boundary $g$** is less accurate 
due to finite size effects. Models with $2n$ spins are used to obtain $g^*(2n)$, which 
is then extrapolated; models of $4n$ spins are needed for $g$**$(4n)$ in order to decouple 
to HAFs with an even number of spins rather than two radicals with $S = 1/2$ ground states. 
Even for large $N$, weak coupling 
between open shell radicals with degenerate GS is quite different from weak coupling of closed 
shell systems with nondegenerate GS. Indeed, Fadeev and Takhtajan \cite{r26} have pointed out 
that the HAF with odd $N$ and states with half-integer $S$ has unexpected and unexplored 
features. Different approaches are required for $1/g \ll 1$, including field theories designed 
for weakly coupled HAFs.  
\section{Discussion}
The defining features of BOW phases are a doubly degenerate GS, 
broken inversion symmetry at sites and finite magnetic gap $E_m$ 
that opens slowly at a Kosterlitz-Thouless transition. The elementary excitations in 
spin chains are $s = 1/2$ solitons centered on sites in opposite sublattices. 
The frustrated spin chains $H(g,\alpha)$ in Eq. \ref{eq1} do not meet these 
signatures at large $g$. To be sure, we cannot discriminate between exponentially 
small and zero $E_m$. But the energy spectrum at $1/g = 0$ is just the HAF spectrum. 
The GS is not degenerate at $1/g = 0$ and it is unlikely that arbitrarily small $1/g$ 
will place $^1|T\rangle|T \rangle$ below $E_m$. Yet that is minimally required 
for GS degeneracy at any $1/g > 0$. Conversely, the lowest triplet state at 
small $g$ is not degenerate while the lowest triplet at large $g$ is doubly 
degenerate, $|T \rangle |G\rangle$ or $|G \rangle |T\rangle$ with both 
unpaired spins largely confined to one sublattice.\\

The principal goal of the present study is the identification 
of a decoupled phase at large frustration g that is distinct from the 
BOW phase of the $J_1-J_2$ model at intermediate $g$. GS is not 
degenerate in the decoupled phase; inversion symmetry is not broken; 
the lowest triplet is doubly degenerate. The magnetic gap $E_m$ is zero within 
numerical accuracy for $g > g$** and strictly so in the model with uniform 
exchange. The properties of the decoupled phase at $g > g_{10}$ in 
Table \ref{tb2} or in Figs. [2-6] are already 
close to the $1/g = 0$ limit of decoupled HAFs on sublattices.\\ 

 BOW phases occur in spin chains \cite{r5,r6,r7,r12} 
with frustrated exchange or in half-filled 1D Hubbard models nearest-neighbor \cite{r11} 
or long-range \cite{r27} Coulomb interactions. Moderate interactions are required 
to avoid a first order transition that for $H(g,\alpha)$ is related to 
LRO at $g = 0$. Models without LRO have continuous transitions from 
spin liquid to BOW to decoupled at $g$*$(\alpha)$ and $g$**$(\alpha)$, 
respectively. As shown in Fig. \ref{fig7}, all spin chains have a
decoupled phase at large $g$. By contrast, Sandvik’s phase diagram \cite{r10} 
at large $g$ distinguishes between models with and 
without LRO and does not terminate the BOW phase.  \\
 
Frustration ensures the existence of the correlated phases with variable 
sublattice spin $\langle S^2_A \rangle$. The model with uniform 
exchange has a first order transition at $g_c = 1/4ln2$ from $S^2_A = n(n +1)$ to $S^2_A = 2$. 
The divergence of $\langle S^2_A \rangle /4n$ is linear or logarithmic, 
respectively, at $g = 0$ in spin chains with and without LRO. The $\alpha = 1$ 
model has a first order transition from a GS with LRO to the decoupled phase. The 
$J_1 - J_2$ model has $\langle S^2_A \rangle = 3n/2$ at $g_{MG} =1/2$ in 
the BOW phase and continuous transitions to a spin liquid phase $g < g$* 
and the decoupled phase at $g > g$**. The BOW phase narrows in models 
with intermediate $\alpha$ and terminates for $\alpha < \alpha_c \approx 1.8$. 
Accurate phase boundaries pose major challenges for all models and 
especially for systems with continuous transitions and a BOW phase.

{\bf Acknowledgments:} ZGS thanks A.W. Sandvik for a stimulating discussion of spin chains. 
We thank the National Science Foundation for partial support of this work through 
the Princeton MRSEC (DMR-0819860). MK thanks  DST India for partial financial 
support of this work.

\end{document}